\documentclass[12pt,preprint]{aastex}

\newcommand{\tts}{T\,Tau\,S}
\newcommand{\ttb}{T\,Tau\,Sb}
\newcommand{\tta}{T\,Tau\,Sa}
\newcommand{\ttn}{T\,Tau\,N}
\newcommand{\brg}{Br$\gamma$}

\shorttitle{Resolved Spectroscopy of \tts}
\shortauthors{Duch\^ene et al.}

\begin{document}

\title{Resolved Near-Infrared Spectroscopy of the Mysterious \\
Pre-Main Sequence Binary System \tts}

\author{G. Duch\^ene, A. M. Ghez \& C. McCabe}
\affil{Division of Astronomy and Astrophysics, UCLA, Los Angeles, CA
90095-1562}
\email{duchene@astro.ucla.edu}

%\altaffiltext{1}{Visiting Astronomer, W. M. Keck Observatory.}

\begin{abstract}
We obtained new near-infrared images of the prototypical pre-main
sequence triple system T\,Tau, as well as the first resolved
medium-resolution spectra of the close pair {\tts}. At the time of our
observations, the tight binary had a 13\,AU projected separation and
showed significant motion since its discovery, three years before. The
orbit cannot be strongly constrained yet, but the observed motion of
{\ttb} with respect to {\tta} suggests that the system is at least as
massive as {\ttn} itself. This may indicate that {\ttn} is not the
most massive star in the system. The spectrum of {\tta}, which is
totally featureless except for a strong {\brg} emission line,
identifies this component with the ``infrared companion'', whose exact
nature remains obscure but may be the consequence of it being the most
massive component of the system. Contrasting sharply with {\tta}, the
spectrum of {\ttb} shows numerous photospheric features consistent
with an early-M spectral type. The presence of a strong {\brg}
emission line and of a significant veiling continuum classifies this
object as a deeply embedded T\,Tauri star.  From these observations,
we conclude that both components of {\tts} are embedded in their own
dense circumstellar cocoon of material, which are probably fed by a
much more extended structure.
\end{abstract}


%________________________________________________________________

\section{Introduction}

T\,Tau has long been identified as one of the brightest low-mass
pre-main sequence objects and, as such, has been considered as the
prototype for the T\,Tauri class of objects \citep{joy45}. Its notable
properties include a significant infrared continuum
excess thought to be related to the presence of an accretion disk
around this million year-old object \citep{bertout88}, a bright
nebulosity caused by photon scattering on the inner wall of a cavity
emptied by a strong polar outflow \citep{stapelfeldt98}, a
substantial circumstellar disk resolved through interferometric
imaging at radio wavelengths
\citep{akeson98} and a limited interstellar extinction
\citep[$A_V\sim1.5\,$mag,][]{ghez91}. Early near-infrared high-angular
resolution 
observations revealed a close \citep[0\farcs7, or 100\,AU at
$d=140\,$pc,][]{elias78} companion in this system \citep{dyck82}, which
raised the possibility that T\,Tau is not so prototypical of its
class. From subsequent various high-angular resolution surveys
\citep{ghez93,leinert93,simon95} however, it rapidly became clear that
binarity is not a rare property, re-establishing T\,Tau as a prototype
of pre-main sequence solar-type stars once again.

The companion to T\,Tau, known as \tts, is a very peculiar one. While
it is undetected in visible images obtained first by speckle
interferometry \citep{gorham92} and more recently by HST/WFPC2
\citep{stapelfeldt98}, implying a huge flux ratio at those wavelengths
($\Delta V>9.5$), its bolometric luminosity is about twice as large as
that of {\ttn}. Indeed, {\tts} dominates the system flux at
wavelengths $\gtrsim2\,\mu$m and its spectral energy distribution
peaks around 3\,$\mu$m, a significantly longer wavelength than normal
for T\,Tauri stars \citep{ghez91}. Near- to mid-infrared monitoring
has further revealed that {\tts} is strongly variable, up to 2~mag
over timescales of a few months \citep{ghez91}. Finally, its
near-infrared spectrum is featureless around 2\,$\mu$m at a resolution
of $R\sim760$, with the exception of a large {\brg} emission line and
a much weaker H$_2$ line \citep{beck01}. Together with a handful of
other companions to T\,Tauri stars, {\tts} has been classified as an
``infrared companion'' (IRC). As reviewed by \cite{koresko97}, the
peculiar properties of these companions has generated a wide range of
theories regarding their nature, which include an embedded
intermediate- to high-mass protostar, a planetary object embedded in
the disk surrounding their optical companion or a strongly accreting
FU\,Ori-like object among others possibilities. The very nature of
{\tts} and other IRCs is still debated as none of these explanations
fit all the observational data.

Increasing the complexity of the T\,Tau system, \cite{koresko00}
identified a very close companion to {\tts} through speckle
holography. In the discovery images, the companion was projected a
mere 7\,AU (0\farcs05) away from the IRC. Such a small separation
raised new questions about the {\tts} system: are both components
classifiable as IRCs? If embedded, are they inside the same cocoon of
material? Can the additional companion be used to clarify the nature
of IRCs? While previous high-angular resolution observations of this
system never resolved the IRC \citep[e.g.,][]{simon96}, follow-up
observations by \cite{kohler00} confirmed the presence of a close
companion. However, they noted significant changes in the binary
separation and position angle, which cannot be accounted for by
measurement uncertainties, but rather suggests that the two objects
are bound with an orbital period on the order of 10--20\,yrs.

In this paper, we present adaptive optics images obtained with
the Keck telescope as well as the first spatially resolved $K$-band
spectra of the {\tts} system. Our observations and data reduction
procedures are described in \S\,\ref{sec:obs-red} and our main
results are presented in \S\,\ref{sec:results}. Implications on the
nature of the IRC and its companion are presented in
\S\,\ref{sec:discus}. Finally, \S\,\ref{sec:conclu} summarizes
our findings.

%________________________________________________________________

\section{Observations}
\label{sec:obs-red}

The observations presented in this paper were obtained on 2000
November 19 on the 10\,m Keck II telescope with the facility
near-infrared spectrometer NIRSPEC \citep{mclean00} installed behind
the adaptive optics system \citep{wizinowich00}. {\ttn} ($R=9.2$) was
used as a natural guide star for the adaptive optics system and images
of the whole T\,Tau system, including {\ttn}, were obtained using
SCAM, the slit-viewing camera, with both $H$\footnote{The filter we
used, ``NIRSPEC-3'', is a customized filter specific to the
instrument. Its central wavelength ($\lambda_c=1.625\,\mu$m) and
bandwith ($\Delta\lambda=0.36\,\mu$m) are close enough to the standard
broadband $H$ filter ($\lambda_c=1.630\,\mu$m,
$\Delta\lambda=0.31\,\mu$m) that we do not expect magnitudes
differences larger than a few hundredths. Throughout our paper, we
therefore refer to this filter as the $H$ filter.} and $K$ broad-band
filters at four dithered positions over the detector to correct for
bad pixels and flat-field variations. Total integration times were 40
seconds with both $H$ and $K$ filters and we achieved Strehl ratios of
24\,\% at $H$ and 30\,\% at $K$. The resolution of the images,
measured as the FWHM of a gaussian fit to the radial profile of the
unresolved star {\ttn}, are respectively 40 and 50\,mas at $H$ and
$K$. Both components of {\tts} appear unresolved at these resolutions
in our images.

Long slit $K$-band medium-resolution spectra of the {\tts} components
were obtained by aligning the 0\farcs036-wide slit along the binary
position angle; four spectra were obtained at various locations behind
the slit, amounting to a total exposure time of 8 minutes. The spectra
extend from 2.0$\,\mu$m to 2.4$\,\mu$m with a 4.3\,\AA/pixel scale
and have a spectral resolution of $R\sim3500$ as measured from the
FWHM of various unresolved arc lines. Arc lamps were observed for
wavelength calibration and a halogen lamp image was used for
estimating the pixel-to-pixel response of the detector. A spectrum of
HR\,1388, an A7 dwarf, was obtained with the same set-up immediately
before {\tts} at a comparable airmass to correct for the numerous
atmospheric absorption lines and bands. To remove the strong {\brg}
absorption line for this A-type star, we also observed HD\,1520, an F8
star, whose spectra can be compared to the well-known solar spectra
(G2 spectral type) to isolate telluric absorption lines in the
vicinity of the hydrogen line, using the method described by
\cite{hanson96}. We also observed a few late-type stars of known
spectral type with the same set-up to be used as templates. These
objects are 61\,Cyg\,A, Gl\,123 and HD\,1326, whose spectral types are
K5, M0, and M2 respectively.

All data reduction procedures were performed using IRAF\footnote{IRAF
is distributed by the National Optical Astronomy Observatories, which
is operated by the Association of Universities for Research in
Astronomy, Inc., under contract to the National Science Foundation}
for both imaging and spectroscopic data. Accurate astrometry and
relative photometry were extracted by point spread function fitting,
using {\ttn} as an unresolved source, and the standard deviations
between the four dithered positions gave us an estimate of the
uncertainties for both filters. The astrometry was then combined with
the absolute orientation of the detector on the sky, known to
$\pm0\fdg1$ or so, and its known plate scale (17.2$\pm$0.1\,mas/pixel,
D. Le Mignant, priv. comm.). The sharper images at shorter wavelenghs,
together with the variation of flux ratio with wavelength in the
{\tts} system, are such that the binary is more easily resolved in the
$H$-band images and, although both filters yielded compatible
measurements, the astrometric binary parameters given in this paper
were extracted from the $H$-band image only.

The pipeline developed to reduce the spectroscopic data performs the
following usual reduction steps: sky subtraction, distortion
corrections, flat-fielding, bad pixels and cosmic rays removal. To
extract the spectra of both components, a gaussian profile, which
represents the core of the stellar profiles very well, was fit to both
stars along the spatial direction for each of the 1024 spectral slices
of the reduced spectrum; the width of the gaussian was forced to be
the same for both stars in that process. A Lorentzian profile, which
provides a better fit of the wings of the point spread function, was
also used in a separate analysis of the same reduced spectra, but
resulted in noiser, though similar, spectra; this is due to the
relatively high Strehl ratios in our images, which ensure that most of
the flux is contained in the stellar cores. In the following, we thus
adopt the spectra extracted using a gaussian profile. We estimate that
contamination introduced by the wing of the primary's point spread
function on the spectrum of the secondary amounts to less than 10\,\%
of the latter at any wavelength based on the observed radial profile.

The dominant sources of noise in the final spectra are the gaussian
fitting process and the atmospheric absorption correction. Estimates
for the signal-to-noise ratios can be obtained by studying line-free
sections of the stellar spectra; the signal-to-noise ratio in the
final spectra is about 60--80 per pixel\footnote{The spectral
resolution element is about 50\,\% larger than the pixel size.} for
both stars in the spectral range 2.08--2.38\,$\mu$m. Outside this
range, the quality of the spectra drops sharply because of strong
atmospheric absorption features, poor filter transmission, and
detector response.

%________________________________________________________________

\section{Results}
\label{sec:results}

%___    ___    ___    ___    ___    ___    ___    ___    ___    ___

\subsection{Astrometry and relative photometry}

\begin{figure}[t]
\begin{center}
\plotone{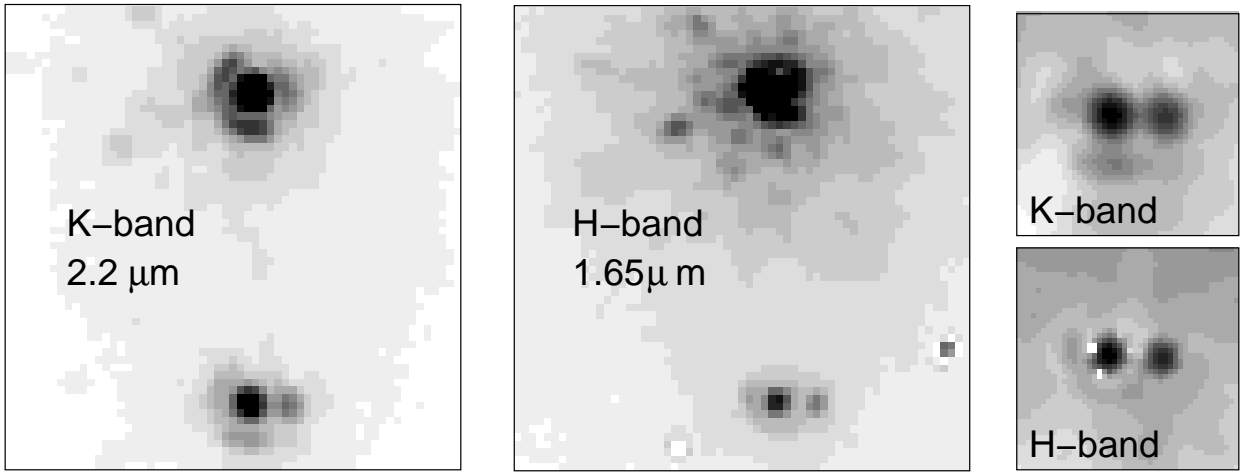}
\caption{{\it Left and center}: $K$- and $H$-band images of the
whole T\,Tau system. A square-root strech was used to allow a proper
identification of all components of the system; {\ttn} dominates the
system at both wavelengths. North is up, and East to the left. Both
images are approximately 1{\arcsec} on each side and {\ttb} is located
just West of {\tta}. In spite of the increased brightness of both
{\tts} components at longer wavelength, the two are best resolved in
the $H$-band image. {\it Right}: Images of {\tts} lightly deconvolved
with a Lucy-Richardson algorithm using {\ttn} as the empirical point
spread function. The images have been resampled by a factor of two in
both directions and the image size is about 0\farcs4.}
\label{fig:images}
\end{center}
\end{figure}

Our final $H$- and $K$-band images are presented in
Fig.\,\ref{fig:images}, while the astrometry and relative photometry
extracted through point spread function fitting are summarized in
Table\,\ref{tab:res1}. {\tts} is clearly resolved at both
wavelengths. To confirm the detection of the two components and to
confirm our astrometric and photometric measurements, we applied a
light deconvolution process with the Richardson-Lucy algorithm
implemented in IRAF. {\ttn} was used as an unresolved point spread
function and we ran 10 iterations after having resampled the images by
a factor of two in each direction. The deconvolved images are also
shown in Fig.\,\ref{fig:images}. The two components are nicely
recovered and the quantitative information presented in
Table\,\ref{tab:res1} agree very well with the deconvolved images.
The location of {\ttb} with respect to {\tta}, which is significantly
different from previous measurements, confirms the relative motion of
the two stars reported by \cite{kohler00}. At the time of our
observations, the projected separation of the binary was
$12.9\pm0.4$\,AU. The nature of the observed motion is discussed in
\S\,\ref{sec:discus}.

%\placetable{tab:res1}

%\placefigure{fig:images}

The flux ratio between the two components of {\tts} has also changed
over the last few years. \cite{koresko00} measured $\Delta K_{Sa-b} =
2.61\pm0.24$ in December 1997 and \cite{kohler00} found $\Delta
H_{Sa-b} = 1.47\pm0.05$ in February 2000, which both differ from our
observations by $\gtrsim3\,\sigma$. In spite of the lack of absolute
photometry, the variable source can be deduced from the $K$-band
datasets. The $K$-band flux ratio between {\ttn} and {\tta} is
identical, to within statistical uncertainties, and the $K$-band flux
of {\ttn} has been almost constant over the last ten years
\citep{kobayashi94,beck_simon01}. Therefore, we conclude that {\ttb}
is itself variable and brightened between the two epochs while {\tta}
remained constant. Our measurements also show that {\tts} as a whole
was close to its bright state at the time of our observations, since
the magnitude difference between {\ttn} and the unresolved {\tts}
system appears to vary from $\sim0.5$ to $\sim2.5\,$mag
\citep{ghez91,roddier00}. Furthermore, if {\tta} always dominates its
companion at $K$, the known variability of the unresolved {\tts}
system indicates that both components are variable.

%___    ___    ___    ___    ___    ___    ___    ___    ___    ___

\subsection{{\tta} spectrum}
\label{subsec:spectta}

The spectrum of {\tta} is in general featureless apart from a
prominent {\brg} emission line which has an equivalent width of
3.8$\pm$0.2\,\AA\ (see Fig.\,\ref{fig:spectra} and
Table\,\ref{tab:res2}). Similar equivalent widths for that line were
reported by \cite{herbst96} and \cite{beck01}, although we failed to
detect the 2.12\,$\mu$m H$_2$ emission line that was detected with an
equivalent width of 3.1$\pm$1.4\,{\AA} by the latter group; our
$3\,\sigma$ upper limit on its equivalent width is 0.5\,\AA. A likely
explanation for that discrepancy is that the emission could arise from
the gas surrounding the star, as already suggested by \cite{beck01};
their observations were performed through a 0\farcs8-slit which is
much wider than ours, and the surrounding nebulosity has already been
detected in this H$_2$ line \citep{vanlangevelde94}. Otherwise, this
line could be strongly variable in {\tta} or the previous detection,
which is only marginally significant, is spurious.

%\placetable{tab:res2}

%___    ___    ___    ___    ___    ___    ___    ___    ___    ___

\subsection{{\ttb} spectrum}

\begin{figure*}[t]
\plotone{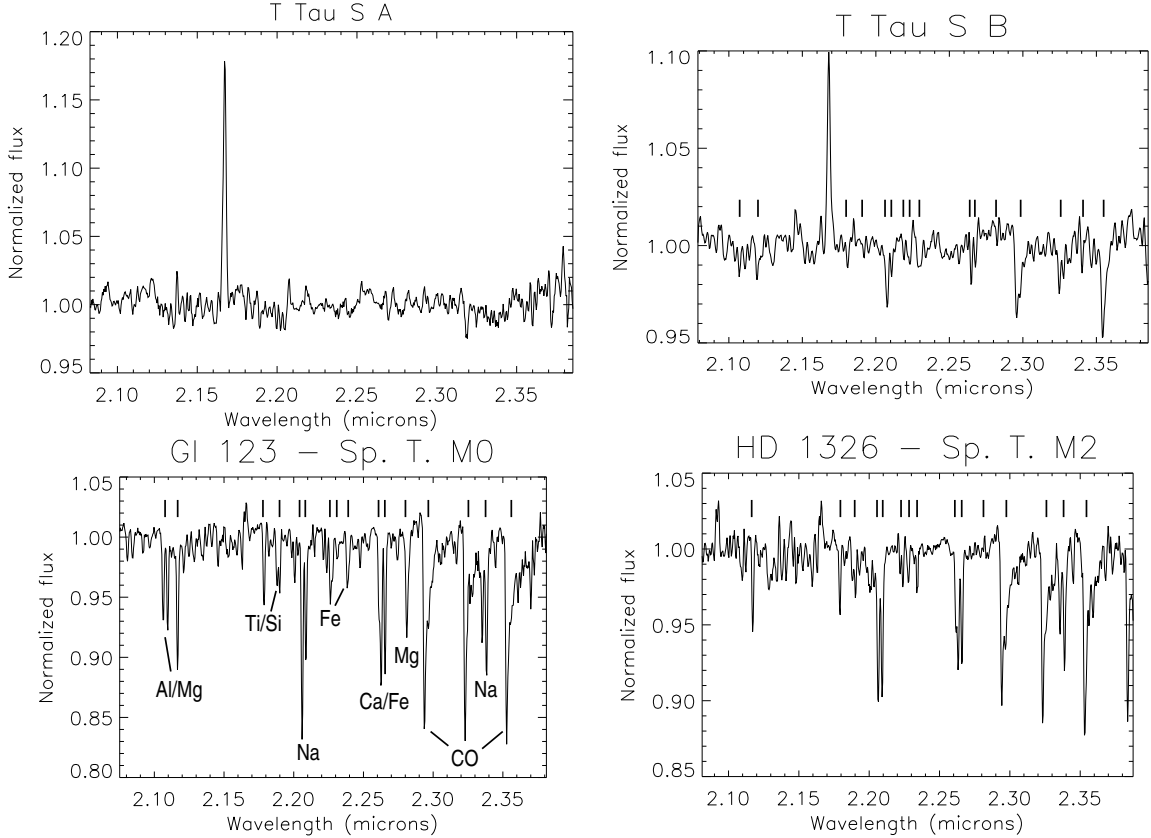}
\caption{$K$-band spectra of {\tta} (upper left) and {\ttb} (upper
right), compared to M0 and M2 dwarfs observed during the same run
(lower left and right spectra, respectively). All spectra were
convolved to a final resolution of $R\sim1500$ to increase the
signal-to-noise ratio and normalized to a low-order polynomial
continuum. Note the different vertical scale for each spectrum. The
{\tta} spectrum was corrected for atmospheric absorption using
HR\,1388 as a calibrator, while the {\ttb} spectrum was divided by the
(featureless) spectrum of {\tta} (see text). All features used for
spectra typing are marked in the spectra of {\ttb} and the dwarf
standards and are identified in the spectrum of Gl\,123 for
comparison.}
\label{fig:spectra}
\end{figure*}

Several features, including the $^{12}$CO ($\Delta\nu=2$) bands at
$\sim2.3\,\mu$m and beyond and the Na doublet at 2.2$\,\mu$m, can be
detected in the spectrum of {\ttb} when using HR\,1388 as the
atmospheric calibrator. However, a slight mismatch in airmass (by
about 0.08) between this calibrator and {\tts} and the time dependence
of the atmosphere impede a perfect correction. To more accurately
remove the telluric features from the spectrum of {\ttb}, we used the
spectrum of {\tta}, which was observed simultaneously and was found to
be featureless with the exception of {\brg} (see
\S\,\ref{subsec:spectta}). The {\brg} line appears ``in emission'' in
the spectrum of {\ttb} shown in Fig.\,\ref{fig:spectra} because the
equivalent width of this line is larger in the spectrum of the
secondary, but its actual strength cannot be directly measured from
this plot. Therefore, the equivalent width of the hydrogen line of
{\ttb} was extracted from the spectrum using HR\,1388 as a calibrator,
while the discussion of all other parts of the spectrum refers to that
obtained using {\tta} as a calibrator. No emission is detected at the
wavelength of the 2.12\,$\mu$m H$_2$ line, with a $3\,\sigma$ upper
limit of 0.5\,{\AA} for its equivalent width.

%\placefigure{fig:spectra}

The spectrum of {\ttb} contains a number of significant features that
appear in all four individual spectra. Among these, sixteen atomic and
molecular features are confidently identified (and labelled in
Fig.\,\ref{fig:spectra}), whose relative strength allow us to estimate
the spectral type of {\ttb}. This can be done by comparing the
spectrum we obtained to spectral atlases found in the literature or to
the standard stars observed during the same run, which is preferred
since the instrumental set-up matches perfectly that used for
{\tts}. As can be seen from Fig.\,\ref{fig:spectra}, the spectrum of
{\ttb} seems very similar to those of the two early-M stars Gl\,123
and HD\,1326. The K5 star (61\,Cyg\,A) we also observed shows more and
stronger Si and Ti features with comparison to the other spectral
lines and does not match well the spectrum of {\ttb}. The latter shows
stronger Al and Mg features at $\sim2.11\,\mu$m than HD\,1326 (M2),
but a much weaker Mg line at 2.28$\,\mu$m than Gl\,123 (M0). On the
other hand, the ratio of the two lines of the Na doublet at
2.2$\,\mu$m appears very similar in {\ttb} and Gl\,123 while that of
HD\,1326 is significantly different. Although a very accurate estimate
of the spectral type would require a higher signal-to-noise spectrum
as well as more spectral type standards, we conclude that {\ttb} has a
spectral type M1 with an uncertainty of about one subclass. A similar
comparison of the spectrum of {\ttb} to spectral atlases in the
literature \citep{kleinman_hall86,ali95,wallace_hinkle97} suggests a
spectral type later than K6 and earlier than M2--M3, consistent with
the result presented above.

It is interesting to note that all the features detected in the
spectrum of {\ttb} appear much weaker than in the early-M spectral
standards observed at the same resolution. The equivalent width of the
strongest features (CO bands, Na doublet, Ca triplet at 2.26$\,\mu$m,
Al line at 2.117$\,\mu$m) across the whole $K$-band are indeed $\sim3$
times smaller than that of the spectral standards. This is quite
unlikely to be the cause of a mismatch in luminosity class, since
\cite{kleinman_hall86} have shown that only the CO bands are very
sensitive to this parameter. Furthermore, although T\,Tauri stars are
frequently thought of as subgiants, the latter have stronger, and not
weaker, CO bands. It thus seems that the spectrum of {\ttb} contains,
in addition to its M1 photospheric component, a roughly flat continuum
throughout the 2.1--2.35$\,\mu$m range. This $K$-band excess must
represent about 200\,\% of the photospheric flux to account for the
observed weakness of the various features. This is much larger than
the 10\,\% upper limit we placed on the contamination from the
featureless spectrum of the primary, and we conclude that it is
intrinsic to the object.

%________________________________________________________________

\section{Discussion}
\label{sec:discus}

%___    ___    ___    ___    ___    ___    ___    ___    ___    ___

\subsection{{\ttb} : A normal active T\,Tauri star}

The unresolved appearance and the photospheric features in the
spectrum of {\ttb} both point to a stellar nature for this
object. Although it cannot be definitely excluded that we are seeing
scattered light off a compact, dense dust clump in the environment of
{\tta} whose intrinsic spectrum would then be that of a M1 dwarf, the
presence of a strong emission at {\brg} in both components with
different equivalent width, makes this scenario unlikely. Overall,
{\ttb} appears to be a regular T\,Tauri star, showing a photospheric
spectrum and, additionally, strong accretion signatures such as a
large {\brg} emission, and significant veiling
($r_K\approx2$). Although significant emission in the hydrogen line
can arise from chromospheric emission in M-type dwarfs
\citep[e.g.,][]{martin98}, we note that the equivalent width of the
{\brg} line is very strong in {\ttb}: only $\sim15\,\%$ of the
classical T\,Tauri stars in Taurus studied by \cite{muzerolle98} show
larger equivalent widths. Therefore we believe that emission in this
line results from the usual T\,Tauri activity mechanisms.

The only unusual property of this object might well be its apparent
faintness, which suggests that the extinction along the line of sight
to {\ttb} is fairly large. Its $H-K$ color index, obtained by
combining our relative photometry with the relatively stable absolute
photometry for {\ttn} from \citeauthor{kenyon_hartmann95}
(\citeyear{kenyon_hartmann95}; $K=5.56$ and $H=6.36$), is about
1.9$\pm$0.3\,mag. This is much redder than expected for a M1 dwarf
($H-K\sim0.2$\,mag) and probably is the result of a combination of
extinction and wavelength-dependent infrared excess related to the
accretion phenomenon. Since its spectrum reveals photospheric
features, it is possible to get a lower limit to the actual extinction
towards this object by assuming that the continuum excess detected in
the $K$-band does not extend into the $H$-band, where we would detect
only photospheric flux. Since the photospheric flux from the star at
$K$ is only one third of the total flux, the excess-corrected color
index of {\ttb} is $H-K\geq0.7$\,mag. The remaining color excess,
$\Delta(H-K)\geq0.5$\,mag, corresponds to an extinction along the line
of sight of $A_V\geq8$\,mag \citep{rieke_lebofsky85}, which is much
larger than the extinction to {\ttn}. Combining this with the
intrinsic $V-K$ color of a M1 dwarf
\citep[$V-K=3.8$][]{kenyon_hartmann95}, we estimate that the expected
visible magnitude of {\ttb} is $V>19.7$, which explains its
non-detection by \cite{gorham92} and \cite{stapelfeldt98}.

%___    ___    ___    ___    ___    ___    ___    ___    ___    ___

\subsection{Orbital motion within the triple system T\,Tau}
\label{subsec:orbmotion}

\begin{figure}[t]
\plotone{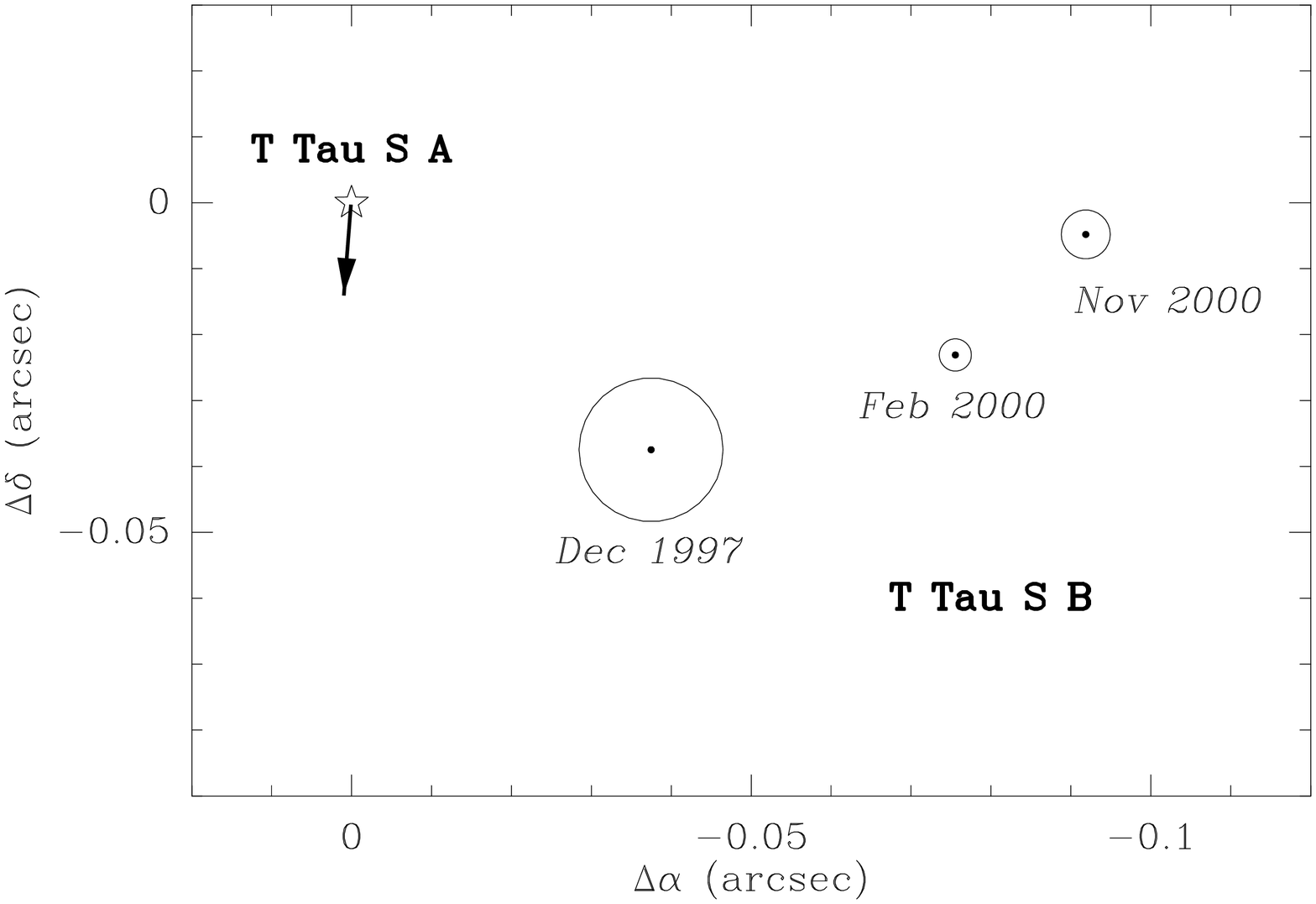}
\caption{Motion of {\ttb} with respect to {\tta}. The large circles
represent the 1$\,\sigma$ uncertainties of the three detections of the
binary to date \citep[][ and this paper]{koresko00,kohler00}. The
arrow represents the expected proper motion per year of {\tta} with
respect to background stars, which is assumed to be the sum of that of
{\ttn} and the orbital motion of the wide pair (see text for
details).}
\label{fig:orbit}
\end{figure}

As can be seen from Fig.\,\ref{fig:orbit}, {\ttb} is moving very fast
with respect to {\tta} (4.2\,AU\,yr$^{-1}$, or 20\,km\,s$^{-1}$, based
on the latest two measurements). There are two potential explanations
to this: either constant linear motion of a background star or orbital
motion. We first consider the former possibility by comparing the
observed motion to what is expected for a field star and show that
they are unlikely to match each other.

%\placefigure{fig:orbit}

First of all, one can extrapolate the expected location of a chance
alignment background star at the time of the first epoch from the last
two measurements, which are the most accurate. The predicted
separation and position angle for December 1997 are
$0\farcs082\pm0.023$ and $200\pm15\degr$. This is only 1.5$\,\sigma$
different from the measurement of \cite{koresko00}, which is not
enough to exclude a constant linear motion of {\ttb} with respect to
{\tta}. If {\ttb} was a background star, its velocity should be the
sum of the proper motion of {\ttn}, which was measured by
\cite{jones_herbig79}, and the relative motion of the {\ttn}/{\tts}
pair, which has been estimated by \cite{roddier00}. This combined
motion, however, is about twice as slow as the observed motion and its
direction is about forty degrees away. We thus conclude that {\ttb}
is not a background star. This is strongly reinforced by the extremely
low probability, $\sim10^{-7}$ based on the $2.5\times10^{-6}$ per square
arcsecond surface density of objects brighter than $K=9$ estimated
by the 2MASS
survey around T\,Tau, of finding a star as bright as {\ttb} within 0\farcs1 of
{\tta}. Since {\ttb} appears to be a normal T\,Tauri star, it probably
belongs to the Taurus star-forming region, and we should compare its
observed motion to the internal velocity dispersion in the molecular
cloud, which is only 2\,km\,s$^{-1}$
\citep{jones_herbig79,hartmann86}. This is ten times smaller than
what is observed, and it thus seems unlikely that this binary results
from the visual alignment of two unrelated T\,Tauri stars.

Another serious issue for the scenario of a linear, independent motion
of the two {\tts} components is the result of the lunar occultation
observations presented by \cite{simon96}, who failed to detect any
companion to {\tts}. At the time of their observations, in December
1994, {\ttb} should have been located $\sim0\farcs15$ away from {\tta}
if its motion remained constant at the current observed
speed. Although their observations would have failed to detect a
companion in the same configuration as seen by \cite{koresko00}, such
a large system would have been easy to detect unless it was then much
fainter than it is currently (M. Simon, priv. comm.). Together with
the relative speed of {\ttb}, this suggests that the two stars form a
bound system and that, for several years prior to its discovery,
{\ttb} was much closer to {\tta} than it is now.

We now assume that the observed motion is indeed that of a bound
binary system, since it is statistically unlikely that we are
witnessing an unstable triple system at the very moment of its
disruption. Although three measurements are not enough to fit a
complete orbital solution, one can still estimate the total mass of
the system from an order of magnitude
calculation. Fig.\,\ref{fig:orbit} obviously indicates that the
observed motion cannot be a face-on circular orbit: the orbit must be
inclined and/or eccentric. Assuming a 15\,AU-radius inclined circular
orbit, which is one of the simplest possible fits, the observed
velocity implies a total system mass of $5\pm2\,M_\odot$, to be
compared to the $2\,M_\odot$ mass of {\ttn}
\citep[e.g.,][]{koresko97}, and an orbital period of
$25\pm5$\,yrs\footnote{Only the uncertainty on the velocity is
propagated to estimate errors here. We caution that the shape of the
orbit is not constrained enough at this point and that these
uncertainties should not be considered as absolute.}. Such a mass for
the {\tts} pair implies a total mass of $\sim7\,M_\odot$ for T\,Tau, a
factor of two larger than a previous estimate by \cite{roddier00} on
the basis of the wide pair orbital motion (estimated orbital period
$\approx500$\,yr). However, this mismatch is not statistically
significant, as the dependency of orbital velocities to mass is only a
square root effect and the measured velocities are fairly small.
Independent of the details of the orbital motion within {\tts}, we
can reasonably assume that its orbital period is on the order of a few
decades. This implies that T\,Tau, considered as a triple system is
gravitationally stable since the ratio of the orbital periods involved
is at least a factor of ten \citep{eggleton_kiseleva95}.

The presence of a third component in the system might bias attempts to
reconstruct the orbital motion of the wide {\ttn}/{\tts} pair, as the
orbital motion of {\tts} would result in a shift of its photocenter,
which was used in previous studies of the system. However, using
gaussian profiles to model both components of {\tts}, one can show
that the photocenter should not move away from {\tta} by more than
0\farcs005 {\it if} {\ttb} does not get much brighter than it
currently is. Under this assumption, which is unverified in the long
run, the observed motion of the unresolved {\tts} pair mostly relates
to its orbit with {\ttn}.

%___    ___    ___    ___    ___    ___    ___    ___    ___    ___

\subsection{Geometry of the system}
\label{subsec:geom}

Our study confirms that {\tts} is a tight binary system and further
shows that {\ttb} is a strongly extincted early M-type T\,Tauri star,
quite different from {\tta} which is now identified as the actual
IRC. This allows us to revisit the various explanations proposed so
far about the physical nature of the later object. We start by
investigating the geometry of the system, before we move on to the
very nature of the IRC.

One of the most popular theories to account for the properties of
{\tts}, first suggested by \cite{hogerheijde97}, is that it lies
behind the optically thick circumstellar disk surrounding {\ttn}. We
know that this disk is likely seen close to face-on, as both the
analysis of rotation properties of {\ttn} and velocity measurements of
the collimated outflow from that star suggest that its polar axis is
only inclined by about 20$\degr$ to the line of sight
\citep{eisloffel_mundt98}. To prevent disk truncation by gravitational
resonances, the actual separation of the wide pair must be much larger
than the disk itself \citep[cf.,][]{lin_papaloizou93}. This implies that
the {\ttn} disk and {\ttn}--{\tts} orbit are not coplanar, with {\tts}
currently lying well beyond the disk. Therefore, the two components of
{\tts} are likely to suffer roughly from the same extinction. Yet, the
featureless spectrum of {\tta} contrast sharply with the numerous
photospheric lines identified in that of {\ttb}, which suggests that
the disk of {\ttn}, even though it may lie between us and {\tts}, is
not the only factor responsible for the unusual characteristics of the
IRC, unless the disk around {\ttn} is very clumpy on a few AU
scale. We conclude that {\tts} is surrounded by some optically thick
material located much closer to the stars than {\ttn} and its disk
are, as pointed out by the accretion diagnoses found in the spectrum of
{\ttb} which are suggestive of a circumstellar disk.

This analysis suggests that {\tta} is deeply embedded in a substantial
amount of dusty material which reprocesses photons at mid-infrared
wavelengths and annihilates any near-infrared photospheric
feature. Alternatively, {\tta} could be intrinsically featureless,
either because it is an F star, but such strong hydrogen emission
lines and red colors have never been found in young intermediate-mass
stars, or because its spectrum is completely dominated by an
accretion-induced continuum flux. Even if the absence of photospheric
features is not due to photon reprocessing by dust grains, the
extinction towards {\tta} has to be very large to account for its
non-detection in the visible, probably much larger than that towards
{\ttb}\footnote{A visual extinction of $A_V\approx17.4$\,mag has been
derived by \cite{vandenancker99} on the basis of the depth of the
silicate feature towards {\tts}. Given the flux ratio of the tight
binary, this extinction applies to {\tta}.}. Although the {\tts}
binary might be
surrounded by a significant circumbinary envelope, the independent
variability of both components suggests that they both have their own
circumstellar obscuring material, which can be an almost edge-on disk
or a more spherical thick envelope. In any case, this material must be
confined to within a few AUs of each star, as orbital motion of the
close pair would have truncated them long ago otherwise. These
structures must be fed from an envelope surrounding the whole system
in order to avoid complete clearance within a few thousands years: the
extinction-corrected {\brg} flux from {\ttb} only, $F_{Br\gamma}({\rm
T\,Tau\,Sb})\geq 8.5\,10^{-14}$\,erg.cm$^{-2}$.s$^{-1}$, implies an
accretion rate of $\dot{M}\geq10^{-8}\,M_\odot.{\rm yr}^{-1}$ on that
component \citep{muzerolle98}, while the non-detection of thermal
emission from {\tts} by \cite{akeson98} places an upper limit of about
$7\times10^{-4}\,M_\odot$ on the mass of the gaseous material
surrounding the tight binary. The existence of such a vast envelope
was suggested by \cite{hogerheijde97} to fit the mid- to far-infrared
unresolved flux of the system, and it could also account for the past
variability of {\ttn} \citep{beck_simon01}.

Overall, our study does not dramatically modify the geometry depicted
by \cite{solf_bohm99} from their analysis of the two outflows known in
this system, that is {\ttn} seen pole-on in the foreground of {\tts}
which is close to edge-on. However, we believe that the disk of {\ttn}
does not heavily obscure {\tts}. On the other hand, both {\tta} and
{\ttb} possess their own obscuring circumstellar material. The
orientation of the jet emanating from {\tts}, almost in the plane of
the sky, suggests that the star responsible for its launching could be
seen through an almost edge-on disk. It is even possible that both
circumstellar disks are parallel and seen almost edge-on, accounting
for heavy extinction towards both components. In any case, the
presence of accretion diagnostics in the spectra of both components of
{\tts} implies that they are both actively accreting, and hence that
they are very likely surrounded by circumstellar disks. Whether we are
looking at the stars through these disks or at higher inclination
cannot be solved without much higher angular resolution studies
allowed by techniques such as long-baseline interferometry. Finally,
we emphasize that the presence of two stellar objects in the {\tts}
system provides a natural explanation to the non-detection of thermal
flux at millimeter wavelengths \citep{hogerheijde97,akeson98}, as
discussed by \cite{koresko00}; most of the flux at these wavelengths
normally comes from cold material located tens of AU away from the
central star, but such material is likely to have been swept away by
the orbital motion of the binary.

{\tts} has long been known to be the origin of non-thermal radio
emission longwards of $\lambda\sim2$\,cm \citep{phillips93}. This is
thought to be the result of free-free emission either in an active
magnetosphere or in an outflow. Recent high-angular resolution 2\,cm
observations of the T\,Tau system reveals that {\ttb} is likely to be
the source of this emission (K. Johnston {\it et al.} 2001, in
preparation). This may suggest that {\ttb} is the source of the
outflow identified by \cite{bohm_solf94} from this system. On the
other hand, if {\tta} is an accretion-dominated pre-main sequence
object (Fu\,Ori-like), as first suggested by \cite{ghez91}, it is also
likely to drive a significant bipolar jet, as most of the younger,
class I objects do; in that picture {\ttb} could be the source of
magnetopheric free-free radio emission while the optical jet would
originate from {\tta}. It is thus still unclear which of the two
components drives the outflow known to originate in the unresolved
binary.

%___    ___    ___    ___    ___    ___    ___    ___    ___    ___

\subsection{On the nature of the IRC phenomenon}
\label{subsec:IRC}

The spectrum of {\ttb} reveals that it is a normal T\,Tauri star that
is strongly extincted, likely due to the presence of some
circumstellar/binary envelope (see above). On the other hand, the
exact nature of {\tta} remains unclear. The mystery surrounding the
IRC {\tts} has just been shifted towards its brightest component in
the near-infrared. This could be an intermediate-mass star (spectral
type F or so), an accretion-dominated object or a simple T\,Tauri star
suffering a moderate accretion rate. In any case, it is deeply
embedded into some sort of circumstellar material, which is confined
into a volume of a few AU in radius, at most. Although this material
could form a passive circumstellar envelope, a significant accretion
rate on {\tta} is evidenced by the strong {\brg} emission line in its
spectrum. Furthermore, the variability of this object cannot be simply
explained by extinction changes, so that it is has to be related to
some intrinsic variability \citep{vandenancker99}. The most likely
explanation is that the accretion-induced luminosity dominates the
total flux of the object and varies with time, reinforcing the
FU\,Ori-like model.

The total mass estimated for the {\tts} system from the observed
orbital motion (\S\,\ref{subsec:orbmotion}) suggests that {\tta} is
quite massive, as {\ttb} is an early M-type star, and therefore is
likely to be on the order of only $M\approx0.5\,M_\odot$. Although the
mass of {\tta} remains uncertain, it is likely to be at least as
massive as {\ttn}. If it really is the most massive star in the
system, the peculiarity of the IRC could be a direct consequence of
its mass, as already suggested by \cite{koresko97}. Because of its
deeper potential well, the IRC could both retain a much denser
circumstellar envelope than its neighbours and concentrate most of the
feeding material coming from the vast reservoir surrounding the whole
system, thus accounting for its apparent properties. We note however
that, if it is not itself a tight binary consisting of two
$\sim2\,M_\odot$ objects, a mass much larger than $3\,M_\odot$ for
{\tta} is problematic for several reasons: even though T\,Tau is among
the youngest T\,Tauri systems \citep[0.6\,Myr,][]{koresko97}, a
$4\,M_\odot$ is already an early-A or late-B star and it is already
close to or on the main sequence at this age. Its luminosity would
then be in the range 100--1000\,$L_\odot$, which is much larger than
the bolometric luminosity of {\tts} \citep{koresko97}. Furthermore,
such a hot star would create a very strong radiative pressure field
and would likely blow away the surrounding dusty envelope. Either the
system is not coeval, with {\tta} in an earlier evolutionary stage, or
the mass of this star is currently overestimated. A continued
follow-up of the {\tta}/{\ttb} orbit over the next few years will help
to refine the orbital parameters.

%________________________________________________________________

\section{Conclusion}
\label{sec:conclu}

We obtained new $H$ and $K$ images of the T\,Tau triple system with
the adaptive optics system at the 10\,m Keck telescope, as well as the
first resolved medium-resolution ($R\sim3500$) $K$-band spectra of the
two components of {\tts}, identified so far as an IRC to {\ttn}. Our
data reveal that {\ttb} is a heavily extincted, actively accreting M1
pre-main sequence object, i.e. a rather normal T\,Tauri star. On the
other hand, {\tta}, which dominates the system in the near-infrared,
is indeed identified as the IRC: its spectrum shows no feature except
for a significant {\brg} emission line. Unlike previous observations,
we do not detect the 2.12\,$\mu$m H$_2$ emission line in either of the
two components.

The tight {\tts} binary shows significant motion since its first
detection, almost three years prior to our observations. Although it
is not yet inconsistent with a constant linear motion, it seems quite
unlikely that this is because either one of the two objects is a
background object, since they both show strong signs of T\,Tauri-like
activity. Interpreting the observed motion as that of two stars
orbiting each other, and assuming that the orbit is inclined and
circular, we derive rough estimates for the orbital parameters. The
total system mass and orbital periods are on the order of $5\,M_\odot$
and $25$\,yrs, although we emphasize that the uncertainty on the
actual orbit shape is not taken into account here.

Although {\tts} might well be located behind the circumstellar disk of
{\ttn}, we argue here that this is {\it not} the main reason for the
peculiarities of {\tts}, because of the strikingly different
properties of {\tta}, which is now identified as the IRC, and {\ttb}
as well as from an analysis of the optical depth of the circumprimary
disk. Both components of {\tts} are likely embedded in their own
circumstellar material, which could be either an almost edge-on disk
or a dense spheroidal envelope. Because of tidal truncation due to the
tight binary's orbit, this material must be confined to within only a
few AU of the stars, making it too small to be resolved by current
observations. Furthermore, this effect also explains the non-detection
of thermal flux in the radio domain, as most of the material located
in the corresponding range of distance from the stars has been swept
away. In any event, the subsistence of dense envelopes on a few AU
scale around both components of {\tts} for several $10^5$\,yrs
requires that they are being replenished by material coming from a
larger scale reservoir such as a vast envelope surrounding the whole
system. The presence of such a structure was already suspected from
the analysis of the spectral energy distribution from the system.

Following the orbit of {\tts} for a few more years will bring a
definitive conclusion regarding the possibility that the system is
simply a projected, unphysical one and provide a much better estimate
of the orbit shape and parameters if the two stars are
bound. Furthermore, high-spectral resolution data will yield radial
velocities, which are very powerful tools to better constrain the
orbit, and potentially reveal for the first time some photospheric
features, which would greatly help understanding the nature of this
peculiar object. We also note that both components of {\tts} are prime
targets for the upcoming long-baseline interferometry experiments, as
the environment of these stars can only be resolved through these
techniques.

%________________________________________________________________

\acknowledgments

Data presented herein were obtained at the W. M. Keck Observatory,
which is operated as a scientific partnership among the California
Institute of Technology, the University of California and the
National Aeronautics and Space Administration. The Observatory was
made possible by the generous financial support of the W. M. Keck
Foundation. The Observatory staff supported our observations very
efficiently and we wish to thank them, more particularly Joel Aycock,
David Le Mignant and David Sprayberry. We also thank Bruce Macintosh
for his help during the observations and Franck Marchis for providing
us with synthetic point spread functions used to estimate Strehl
ratios in our images. We ackowledge our anonymous referee for his
prompt and helpful report, as well as Michal Simon for extensive
discussions about {\tts} and valuable comments on an early version of
this paper. This work has been supported in part by the National
Science Foundation Science and Technology Center for Adaptive Optics,
managed by the University of California at Santa Cruz under
cooperative agreement No. AST-9876783, by NASA's Origins of Solar
System program grant No. NAG-6975 and by the Packard Foundation. This
research makes use of the SIMBAD database, operated at CDS,
Strasbourg, France, and of data products from the Two Micron All Sky
Survey, which is a joint project of the University of Massachusetts
and the Infrared Processing and Analysis Center/California Institute
of Technology, funded by the National Aeronautics and Space
Administration and the National Science Foundation. The authors wish to
extend special thanks to those of Hawaiian ancestry on whose sacred
mountains we are privileged to be guests. Without their generous
hospitality, none of the observations presented herein would have
been possible.

%________________________________________________________________

%________________________________________________________________

\clearpage

\begin{table*}[t]
\begin{center}
\caption{\label{tab:res1} Astrometric and relative photometric results}
\begin{tabular}{ccccc}
\tableline\tableline
Binary & \multicolumn{2}{c}{Relative astrometry} &
 \multicolumn{2}{c}{Relative photometry} \\ 
& sepn. & P.A.\tablenotemark{a} &  $\Delta K$ (mag) & $\Delta H$ (mag) \\
\tableline
 Sa--Sb & $0\farcs092\pm0\farcs003$ & 267\fdg0$\pm$1\fdg6 &
$1.89\pm0.05$ & $1.17\pm0.02$ \\
 N--Sa & $0\farcs702\pm0\farcs005$ & 179\fdg7$\pm$0\fdg2 &
$1.36\pm0.05$ & $3.17\pm0.02$ \\
 N--S$tot$\tablenotemark{b} & & & $1.18\pm0.03$ & $2.87\pm0.02$ \\
\tableline
\end{tabular}
\tablenotetext{a}{Position angles are measured Eastwards from North.}
\tablenotetext{b}{S$tot$ stands for the unresolved {\tts} pair.}
\end{center}
\end{table*}

\begin{table}[t]
\begin{center}
\caption{\label{tab:res2} Spectral results\tablenotemark{a}}
\begin{tabular}{cccc}
\tableline\tableline
Object & S.T. & EW {\brg} (\AA) & EW H$_2$ (\AA) \\
\tableline
 Sa & -- & 3.8$\pm$0.2 & $<0.5$ \\
 Sb & M1 & 6.5$\pm$0.2 & $<0.5$ \\
 N & K1\tablenotemark{b} & 3.0$\pm$0.2\tablenotemark{c} &
$<0.6$\tablenotemark{c} \\
\tableline
\end{tabular}
\tablenotetext{a}{The equivalent widths of the {\brg} and H$_2$
2.12$\,\mu$m lines measure their strength in emission or 3\,$\sigma$
upper limits for those which were not detected.}
\tablenotetext{b}{from \cite{cohen_kuhi79}}
\tablenotetext{c}{from \cite{beck01}}
\end{center}
\end{table}

\begin{thebibliography}{}
\bibitem[Akeson et al.(1998)]{akeson98} Akeson, R.~L., Koerner, D.~W.,
\& Jensen, E.~L.~N.\ 1998, \apj, 505, 358

\bibitem[Ali et al.(1995)]{ali95} Ali, B., Carr, J.~S., Depoy, D.~L.,
Frogel, J.~A., \& Sellgren, K.\ 1995, \aj, 110, 2415

\bibitem[Beck \& Simon(2001)]{beck_simon01} Beck, T.~L.~\& Simon, M.\
2001, \aj, 122, 413

\bibitem[Beck et al.(2001)]{beck01} Beck, T.~L., Prato, L., \& Simon,
M.\ 2001, \apj, 551, 1031

\bibitem[Bertout et al.(1988)]{bertout88} Bertout, C., Basri, G., \&
Bouvier, J.\ 1988, \aap, 330, 350

\bibitem[Bohm \& Solf(1994)]{bohm_solf94} Bohm, K.-H.~\& Solf, J.\
1994, \apj, 430, 277

\bibitem[Cohen \& Kuhi(1979)]{cohen_kuhi79} Cohen, M.~\& Kuhi, L.~V.\
1979, \apjs, 41, 743

\bibitem[Dyck et al.(1982)]{dyck82} Dyck, H.~M., Simon, T., \&
Zuckerman, B.\ 1982, \apjl, 255, L103

\bibitem[Eggleton \& Kiseleva(1995)]{eggleton_kiseleva95} Eggleton,
P.~\& Kiseleva, L.\ 1995, \apj, 455, 640

\bibitem[Eisl\"offel \& Mundt(1998)]{eisloffel_mundt98} Eisl{\"o}ffel,
J.~\& Mundt, R.\ 1998, \aj, 115, 1554

\bibitem[Elias(1978)]{elias78} Elias, J.~H.\ 1978, \apj, 224, 857 

\bibitem[Ghez et al.(1995)]{ghez95} Ghez, A.~M., Weinberger, A.~J.,
Neugebauer, G., Matthews, K., \& McCarthy, D.~W.\ 1995, \aj, 110, 753

\bibitem[Ghez et al.(1993)]{ghez93} Ghez, A.~M., Neugebauer, G., \&
Matthews, K.\ 1993, \aj, 106, 2005

\bibitem[Ghez et al.(1991)]{ghez91} Ghez, A.~M., Neugebauer, G.,
Gorham, P.~W., Haniff, C.~A., Kulkarni, S.~R., Matthews, K., Koresko,
C., \& Beckwith, S.\ 1991, \aj, 102, 2066

\bibitem[Gorham et al.(1992)]{gorham92} Gorham, P.~W., Ghez, A.~M.,
Haniff, C.~A., Kulkarni, S.~R., Matthews, K., \& Neugebauer, G.\ 1992,
\aj, 103, 953

\bibitem[Hanson et al.(1996)]{hanson96} Hanson, M.~M., Conti, P.~S.,
\& Rieke, M.~J.\ 1996, \apjs, 107, 281

\bibitem[Hartmann et al.(1986)]{hartmann86} Hartmann, L., Hewett, R.,
Stahler, S., \& Mathieu, R.~D.\ 1986, \apj, 309, 275

\bibitem[Herbst et al.(1996)]{herbst96} Herbst, T.~M., Beckwith,
S.~V.~W., Glindemann, A., Tacconi-Garman, L.~E., Kroker, H., \&
Krabbe, A.\ 1996, \aj, 111, 2403

\bibitem[Hogerheijde et al.(1997)]{hogerheijde97} Hogerheijde, M.~R.,
van Langevelde, H.~J., Mundy, L.~G., Blake, G.~A., \& van Dishoeck,
E.~F.\ 1997, \apjl, 490, L99

\bibitem[Jones \& Herbig(1979)]{jones_herbig79} Jones, B.~F.~\&
Herbig, G.~H.\ 1979, \aj, 84, 1872

\bibitem[Joy(1945)]{joy45} Joy, A.~H.\ 1945, \apj, 102, 168

\bibitem[Kenyon \& Hartmann(1995)]{kenyon_hartmann95} Kenyon, S.~J.~\&
Hartmann, L.\ 1995, \apjs, 101, 117

\bibitem[Kleinman \& Hall(1986)]{kleinman_hall86} Kleinmann, S.~G.~\& 
Hall, D.~N.~B.\ 1986, \apjs, 62, 501 

\bibitem[Kobayashi et al.(1994)]{kobayashi94} Kobayashi, N., Nagata,
T., Hodapp, K., \& Hora, J.~L.\ 1994, \pasj, 46, L183

\bibitem[K\"ohler et al.(2000)]{kohler00} K\"ohler, R., Kasper, M.~\&
Herbst, T.\ 2000, {\it in} ``Birth and evolution of binary stars'',
poster proceedings of IAU symp. 200, eds. B. Reipurth \& H. Zinnecker,
p. 63

\bibitem[Koresko(2000)]{koresko00} Koresko, C.~D.\ 2000, \apjl, 531,
L147

\bibitem[Koresko et al.(1997)]{koresko97} Koresko, C.~D., Herbst,
T.~M., \& Leinert, Ch.\ 1997, \apj, 480, 741

\bibitem[Leinert et al.(1993)]{leinert93} Leinert, Ch., Zinnecker, H.,
Weitzel, N., Christou, J., Ridgway, S.~T., Jameson, R., Haas, M., \&
Lenzen, R.\ 1993, \aap, 278, 129

\bibitem[Lin \& Papaloizou(1993)]{lin_papaloizou93} Lin, D.~N.~C.~\&
Papaloizou, J.~C.~B.\ 1993, Protostars and Planets III, 749

\bibitem[Mart\'{\i}n(1998)]{martin98} Mart\'{\i}n, E.~L.\ 1998, \aj,
115, 351

\bibitem[McLean et al.(2000)]{mclean00} McLean, I.~S., Graham, J.~R.,
Becklin, E.~E., Figer, D.~F., Larkin, J.~E., Levenson, N.~A., \&
Teplitz, H.~I.\ 2000, \procspie, 4008, 1048

\bibitem[Muzerolle et al.(1998)]{muzerolle98} Muzerolle, J., Hartmann,
L., \& Calvet, N.\ 1998, \aj, 116, 2965

\bibitem[Phillips et al.(1993)]{phillips93} Phillips, R.~B., Lonsdale,
C.~J., \& Feigelson, E.~D.\ 1993, \apjl, 403, L43

\bibitem[Rieke \& Lebofsky(1985)]{rieke_lebofsky85} Rieke, G.~H.~\& 
Lebofsky, M.~J.\ 1985, \apj, 288, 618

\bibitem[Roddier et al.(2000)]{roddier00} Roddier, F., Roddier, C.,
Brandner W., Charissoux, D., V\'eran, J.-P., \& Courbin, F.\ 2000,
{\it in} ``Birth and evolution of binary stars'', poster proceedings
of IAU symp. 200, eds. B. Reipurth \& H. Zinnecker, p. 60
 
\bibitem[Simon et al.(1996)]{simon96} Simon, M., Longmore, A.~J.,
Shure, M.~A., \& Smillie, A.\ 1996, \apjl, 456, L41 

\bibitem[Simon et al.(1995)]{simon95} Simon, M., Ghez, A.~M., Leinert,
Ch., Cassar, L., Chen, W.~P., Howell, R.~R., Jameson, R.~F., Matthews,
K., Neugebauer, G., \& Richichi, A.\ 1995, \apj, 443, 625 

\bibitem[Solf \& B\"ohm(1999)]{solf_bohm99} Solf, J.~\& B{\"o}hm,
K.-H.\ 1999, \apj, 523, 709

\bibitem[Stapelfeldt et al.(1998)]{stapelfeldt98} Stapelfeldt, K.~R.,
Burrows, C.~J., Krist, J.~E., Watson, A.~M., Ballester, G.~E., Clarke,
J.~T., Crisp, D., Evans, R.~W., Gallagher, J.~S.~III, Griffiths,
R.~E., Hester, J.~J., Hoessel, J.~G., Holtzman, J.~A., Mould, J.~R.,
Scowen, P.~A., Trauger, J.~T., Westphal, J.~A.\ 1998, \apj, 508, 736

\bibitem[van Cleve et al.(1994)]{vancleve94} Van Cleve, J.~E.,
Hayward, T.~L., Houck, J.~R., \& Miles, J.\ 1994, American
Astronomical Society Meeting, 184, 4405

\bibitem[van den Ancker et al.(1999)]{vandenancker99} van den Ancker, 
M.~E., Wesselius, P.~R., Tielens, A.~G.~G.~M., van Dishoeck, E.~F., \&
Spinoglio, L.\ 1999, \aap, 348, 877 

\bibitem[van Langevelde, van Dishoeck, van der Werf, \& 
Blake(1994)]{vanlangevelde94} van Langevelde, H.~J., van Dishoeck,
E.~F., van der Werf, P.~P., \& Blake, G.~A.\ 1994, \aap, 287, L25

\bibitem[Wallace \& Hinkle(1997)]{wallace_hinkle97} Wallace, L.~\& 
Hinkle, K.\ 1997, \apjs, 111, 445 

\bibitem[Wizinowich et al.(2000)]{wizinowich00} Wizinowich, P.~L.,
Acton, D.~S., Lai, O., Gathright, J., Lupton, W., \& Stomski, P.~J.\
2000, \procspie, 4007, 2

\end{thebibliography}
\end{document}